\documentclass[11pt]{article}
\usepackage{amssymb}    

  \usepackage[dvips]{graphicx}

\newcommand{\ket}[1]{\left | #1 \right \rangle}
\newcommand{\bra}[1]{\left \langle #1 \right |}
\newcommand{\amp}[2]{\left \langle #1 | #2 \right \rangle}
\newcommand{\proj}[1]{\ket{#1} \bra{#1}}
\newcommand{\tr}{\mbox{Tr} \,}

\def\cb{{\cal B}}

\def\cp{{\cal P}}

\begin{document}
\begin{center}
{\Large\bf An introduction to \\ measurement based quantum
computation\\[4mm]}
Richard Jozsa\\
Department of Computer Science,\\ University of Bristol, Bristol
BS8 1UB, UK.
\end{center}
\begin{abstract}
In the formalism of measurement based quantum computation we start
with a given fixed entangled state of many qubits and perform
computation by applying a sequence of measurements to designated
qubits in designated bases. The choice of basis for later
measurements may depend on earlier measurement outcomes and the
final result of the computation is determined from the classical
data of all the measurement outcomes. This is in contrast to the
more familiar gate array model in which computational steps are
unitary operations, developing a large entangled state prior to
some final measurements for the output. Two principal schemes of
measurement based computation are teleportation quantum
computation (TQC) and the so-called cluster model or one-way
quantum computer (1WQC). We will describe these schemes and show
how they are able to perform universal quantum computation. We
will outline various possible relationships between the models
which serve to clarify their workings. We will also discuss
possible novel computational benefits of the measurement based
models compared to the gate array model, especially issues of
parallelisability of algorithms.

\end{abstract}

\section{Introduction}
Many of the most popular models of quantum computation are direct
quantum generalisations of well known classical constructs. This
includes quantum turing machines, gate arrays and walks. These
models use unitary evolution as the basic mechanism of information
processing and only at the end do we make measurements, converting
quantum information into classical information in order to read
out classical answers. In contrast to unitary evolution,
measurements are irreversibly destructive, involving much loss of
potential information about a quantum state's identity. Thus it is
interesting, and at first sight surprising, that we can perform
universal quantum computation using only measurements as
computational steps \cite{gottch,nielsentqc,leungtqc,rb,rbb}.
These ``measurement based'' models are especially interesting for
fundamental issues: they have no evident classical analogues and
they offer a new perspective on the role of entanglement in
quantum computation. They may also be interesting for experimental
considerations, suggesting a different kind of computer
architecture and offering interesting possibilities for further
issues such as fault tolerance \cite{nielsenft}.

We will discuss two measurement based models. Firstly we'll
consider teleportation quantum computation (TQC). This is based on
the idea of Gottesman and Chuang \cite{gottch} of teleporting
quantum gates, and was developed into a computational model by
Nielsen, Leung and others \cite{nielsentqc,leungtqc}. Secondly we
will consider the so-called ``one way quantum computer'' (1WQC) or
``cluster state computation'' of Raussendorf and Briegel
\cite{rb,rbb}. Then we will discuss various possible relationships
between these two models and finally consider possible
computational benefits of the measurement based formalism as
compared to the quantum gate array model.

The following notations will be frequently used. The Pauli
operators are
\[ I= \left( \begin{array}{rr} 1 & 0 \\ 0 & 1 \end{array} \right)
\hspace{3mm} X= \left( \begin{array}{rr} 0 & 1 \\ 1 & 0
\end{array} \right) \hspace{3mm}
Z= \left( \begin{array}{rr} 1 & 0 \\ 0 & -1 \end{array} \right)
\hspace{3mm} iY=ZX= \left( \begin{array}{rr} 0 & 1 \\ -1 & 0
\end{array} \right) .
\]
$\oplus$ will denote addition modulo 2. The controlled-NOT gate
$CX$ is defined by
\[ CX \ket{i}\ket{j} = \ket{i}\ket{i\oplus j} \hspace{5mm}
i,j=0,1.
\] The controlled phase gate $CZ$ is defined by \[
CZ\ket{i}\ket{j}=(-1)^{ij} \ket{i}\ket{j}.\] In contrast to the
$CX$ gate, $CZ$ is symmetrical in the two input qubits. The
Hadamard operator  is \[ H=\frac{1}{\sqrt{2}}\left(
\begin{array}{rr} 1 & 1 \\ 1 & -1 \end{array} \right) \] and the
Hadamard basis states are
\[ \ket{\pm} = \frac{1}{\sqrt{2}}(\ket{0}\pm \ket{1}). \] The Bell
basis states are
\[ \begin{array}{l}
\ket{B_{00}}= \frac{1}{\sqrt{2}}( \ket{00}+\ket{11}) = I\otimes I
\ket{B_{00}} \\
\ket{B_{01}}= \frac{1}{\sqrt{2}}( \ket{01}+\ket{10}) = X\otimes I
\ket{B_{00}} \\
\ket{B_{10}}= \frac{1}{\sqrt{2}}( \ket{00}-\ket{11}) = Z\otimes I
\ket{B_{00}} \\
\ket{B_{11}}= \frac{1}{\sqrt{2}}( \ket{01}-\ket{10}) = iY\otimes I
\ket{B_{00}} .\\
\end{array} \]
Succinctly we have \[ \ket{B_{cd}}=Z^cX^d\otimes I \ket{B_{00}}.\]
We will also use the maximally entangled state that combines the
Hadamard and standard bases in its Schmidt form:
\[ \ket{H}=\frac{1}{\sqrt{2}}\left(\ket{0}\ket{+}+\ket{1}\ket{-}\right)
= \frac{1}{\sqrt{2}}\left(\ket{+}\ket{0}+\ket{-}\ket{1}\right)=
CZ\ket{+}\ket{+}.\]

\section{Teleportation based quantum computing}

Recall standard teleportation \cite{bbcjpw} as depicted in figure
1 and explained in the caption.

%
\begin{figure}[t] \begin{center}\leavevmode\fbox{\parbox[b][30mm][s]{60mm}{
\vfill\footnotesize \includegraphics{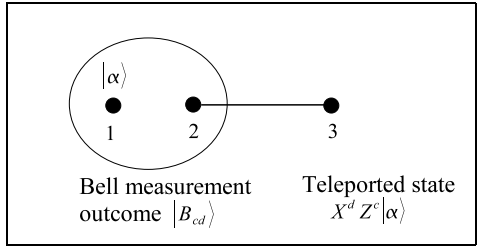} \vfill}}

\caption{\small Standard quantum teleportation. The labels 1,2,3
represent three qubits. 1 is in state $\ket{\alpha}$ and 23 are in
state $\ket{B_{00}}$. A Bell measurement is performed on 12. If
the outcome is $\ket{B_{cd}}$ then qubit 3 acquires the state
$X^dZ^c\ket{\alpha}$. } \label{fig1} \end{center}
\end{figure}

Now introduce the ``rotated Bell basis'' denoted $\cb(U)= \{
\ket{B(U)_{cd}} \}$ where for any 1-qubit gate $U$:
\[ \ket{B(U)_{cd}}= U^\dagger \otimes I \ket{B_{cd}}.\]
A simple calculation shows that if we perform such a rotated Bell
measurement (instead of the standard one in figure 1) then the
teleported state at 3 is $X^dZ^cU\ket{\psi}$ i.e. the gate $U$ has
been applied to $\ket{\psi}$ via this teleportation \cite{gottch}.
A particularly neat and general way of performing this calculation
is the following. For any dimension $d$ consider any maximally
entangled state, written in its Schmidt form as \[ \ket{\phi}=
\frac{1}{\sqrt{d}}\sum_{i=0}^{d-1}\ket{i}\ket{i}. \] Consider the
mathematical projection depicted in figure 2.

\begin{figure}[t] \begin{center} \leavevmode\fbox{\parbox[b][30mm][s]{60mm}{
\vfill\footnotesize \includegraphics{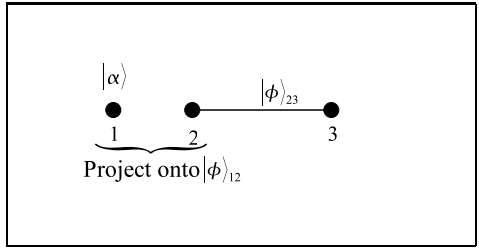}\vfill}}
\caption{\small Projecting the three qubit state
$\ket{\alpha}_1\ket{\phi}_{23}$ onto the maximally entangled state
$\ket{\phi}_{12}$ results in a single qubit state at 3 given by
$\frac{1}{d}\ket{\alpha}_3$ (see lemma 1). }\label{fig2}
\end{center} \end{figure}

\noindent {\bf Lemma 1:}\,\, The projection of
$\ket{\alpha}_1\ket{\phi}_{23} $ onto $\ket{\phi}_{12}$ results in
the state $\frac{1}{d}\ket{\alpha}_3$ at qubit 3.

\noindent {\bf Proof:}\,\, Let $\ket{\alpha}=\sum a_j \ket{j}$.
Then the projection is
\[   \frac{1}{d} (\sum_i \bra{i} \bra{i})
(\sum_{jk} a_j \ket{j}\ket{k}\ket{k})
  = \frac{1}{d} \sum_{ijk} a_j \delta_{ij}\delta_{ik}\ket{k} =
 \frac{1}{d} \sum_k a_k \ket{k}. \hspace{5mm} \Box  \]
 Next introduce the ``rotated $\ket{\phi}$ state'': for any unitary
 operator $U$ in $d$ dimensions define \begin{equation}\label{phiu}
 \ket{\phi (U)}=
 U^\dagger \otimes I \ket{\phi}.\end{equation}

 \noindent {\bf Lemma 2:}\,\,
The projection of $\ket{\alpha}_1\ket{\phi}_{23} $ onto
$\ket{\phi(U)}_{12}$ results in the state
$\frac{1}{d}U\ket{\alpha}_3$ at qubit 3.

\noindent {\bf Proof:}\,\, For any states $\ket{a}$, $\ket{b}$ if
we write $\ket{U^\dagger a}=U^\dagger\ket{a}$ then $\amp{U^\dagger
a}{b}=\amp{a}{Ub}$. Hence lemma 1 with $\ket{\phi}_{12}$ replaced
by $\ket{\phi(U)}_{12}$ immediately gives the result.$\Box$

Now the basic idea is to regard the 12-projections in lemmas 1 and
2 as being outcomes of a projective measurement applied to qubits
1 and 2 i.e. we wish to choose a set of $d^2$ unitaries $U_i$ such
that $\{ \ket{\phi(U_i)} \}$ form an orthonormal basis of the 12
space. For qubits ($d=2$) the set $\{ I,X,Z,XZ \}$ gives standard
teleportation and the set $\{ IU,XU,ZU,XZU \}$ gives the rotated
Bell basis, reproducing our previous procedure of ``gate
teleportation'' to construct $X^dZ^cU\ket{\psi}$ at qubit 3.

Note that lemmas 1 and 2 apply in general dimension $d$ so we can
apply 2-qubit gates such as $CZ$ via measurements by teleportation
in dimension $d=4$. For example using
$\ket{\phi}=\ket{B_{00}}\ket{B_{00}}$ and the 16 operators $
U_{ij}=(P_i\otimes P_j )CZ $ where $P_i$ and $P_j$ range over the
four standard Pauli operators, we can check that $\{
U_{ij}^\dagger\otimes I \ket{\phi} \}$ is an orthonormal set and
the output teleported  state is $(P_i\otimes P_j )CZ\ket{\psi}$
for any input 2-qubit state $\ket{\psi}$.

This method of applying $CZ$ requires a 16 dimensional Bell
measurement but with more subtle means we can achieve the result
with smaller measurements. An illustrative example (taken from
\cite{vc}) is shown in figure 3.

\begin{figure}[t] \begin{center} \leavevmode\fbox{\parbox[b][30mm][s]{60mm}{
\vfill\footnotesize \includegraphics{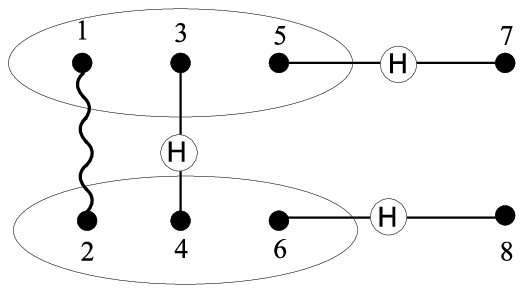}\vfill}}
\caption{\small Application of $CZ$ by teleportation with 8
dimensional measurements. The wiggly line connecting 12 denotes an
input 2-qubit state $\ket{\psi}$. The lines labelled $H$ denote
the maximally entangled state $\ket{H}$. Introduce the 3-qubit
Bell measurement corresponding to the basis $\{
\ket{000}\pm\ket{111}, \ket{001}\pm \ket{110}, \ket{010}\pm
\ket{101}, \ket{100}\pm\ket{011} \}$. If this measurement is
performed on qubits 135 and 246 then it can be verified that
qubits 78 acquire the state $(P_i\otimes P_j)(H\otimes H ) CZ
\ket{\psi}$ where the Pauli operators $P_i$ and $P_j$ depend on
the measurement outcomes.
 }\label{fig3}
\end{center} \end{figure}

\noindent {\bf Remark:} \,\, Returning to $\ket{\phi(U)}$ in eq.
(\ref{phiu}), a set of $d^2$ operators $\{ U_i \}$ will have the
corresponding set $\{ \ket{\phi(U_i)} \}$ being orthonormal iff
$\tr (U_iU^\dagger_j)=\delta_{ij}$ i.e. $\{ U_i\}$ is a so-called
unitary operator basis. For general dimension $d$ many such sets
exist, each corresponding to a teleportation scheme. Werner
\cite{w00} has described a method for constructing a large number
of explicit inequivalent examples. However we can go even further
and choose a set of $n\geq d^2$ operators $U_i$ such that the set
$\{ k_i \proj{\phi(U_i)} \}$ (for some chosen constants $k_i >0$)
form the elements of a (rank 1) positive operator valued measure
(POVM) i.e. we ask that \[ \sum_{i=1}^{n} k_i \proj{\phi(U_i)} =
I_d\otimes I_d.
\]
The case of $n=d^2$ (with all $k_i$'s then necessarily equal)
reproduces projective measurements as above but for $n>d^2$ we
obtain fully valid teleportation schemes in which the Bell
measurement is replaced by a POVM. Even in the qubit case $d=2$
many examples exist and in fact we can have unboundedly large $n$.
(We will not digress here to discuss explicit examples of such
constructions.)$\Box$

Thus in summary so far, if we have a pool of maximally entangled
states we can apply any unitary gate $U$ to any (multi-qubit)
input state $\ket{\psi}$ by measurements alone. A significant
annoyance (and see more about this later) is that we do not get
the exact desired result $U\ket{\psi}$ but instead get
$PU\ket{\psi}$ where $P$ is some Pauli operation (on each qubit)
depending on the measurement outcome. This is the residue of the
randomness of quantum measurement outcomes in our computational
formalism.

To perform universal computation it suffices to be able to apply
gates from any convenient universal set. We introduce  notation
for further gates. $x$ and $z$ rotations are defined by \[
R_x(\theta)=e^{-i\theta X} \hspace{5mm} R_z(\theta)=e^{-i\theta
Z}. \] The following gate will be significant for 1WQC later:
\begin{equation}\label{wth} W(\theta) = \frac{1}{\sqrt{2}} \left(
\begin{array}{cc} 1 & e^{i\theta} \\ 1 & -e^{i\theta} \end{array}
\right) = HP(\theta)\hspace{4mm} {\rm where}\hspace{4mm} P(\theta)
=  \left(
\begin{array}{cc} 1 & 0 \\ 0 & e^{i\theta} \end{array}
\right).
\end{equation}  Any 1-qubit gate $U$ (up to an overall phase) can
be decomposed as (Euler angles):
\[ U=R_x(\zeta)R_z(\eta)R_x(\xi) \hspace{5mm} \mbox{for some
$\xi,\eta,\zeta$} \] and also \cite{vc} as \[
U=W(0)W(\theta_1)W(\theta_2)W(\theta_3) \hspace{5mm} \mbox{for
some $\theta_1,\theta_2, \theta_3$.} \] It is known that $CZ$
together with all 1-qubit operations is a universal set so the
sets $\{ CZ, R_x(\theta), R_z(\phi)\,\, {\rm all}\,\, \theta, \phi
\}$ and $ \{ CZ, W(\theta)\,\, {\rm all}\,\, \theta \}$ are also
universal.

\section{Adaptive measurements}

We wish to perform arbitrary sequences of gates from a universal
set by measurements. For simplicity consider 1-qubit gates: to
perform $\ldots U_3U_2U_1\ket{\psi}$ we successively teleport the
three gates but instead of the desired result we get $\ldots
P_3U_3P_2U_2P_1U_1\ket{\psi}$ where $P_1,P_2,P_3\ldots$ are Pauli
operators depending on the measurement outcomes. To deal with this
awkwardness we introduce a new feature: {\em adaptive choices} of
measurements. Let us assume that all 1-qubit operations are $x$ or
$z$ rotations. The following Pauli ``propagation'' relations are
easily verified:
\[ \begin{array}{c} R_x(\theta)X=XR_x(\theta) \hspace{6mm}
R_x(\theta)Z=ZR_x(-\theta)\\
R_z(\theta)X=XR_z(-\theta)\hspace{6mm}
 R_z(\theta)Z=ZR_z(\theta).
\end{array}
\]
(Alternatively we could use only $W(\theta)$'s with relations
$W(\theta)X=ZW(-\theta)$ and $W(\theta)Z=XW(\theta)$.) Suppose we
want to do $\ldots R_z(\beta)R_x(\alpha)\ket{\psi}$. The first
Bell measurement with outcomes $a,b$ gives
$Z^aX^bR_x(\alpha)\ket{\psi}$. Now commuting $Z^aX^b$ to the left
through $R_z(\beta)$ causes $\beta$ to change to $(-1)^b\beta$. So
if instead of $R_z(\beta)$ we next applied the measurement for
$R_z((-1)^b\beta)$ instead, then \[
R_z((-1)^b\beta)Z^aX^bR_x(\alpha)\ket{\psi}=
Z^aX^bR_z(\beta)R_x(\alpha)\ket{\psi} \] as wanted. This second
measurement with basis adapted to measurement outcome $b$ has an
outcome $c,d$ say, and the state is now
$Z^{a+c}X^{b+d}R_z(\beta)R_x(\alpha)\ket{\psi}$. Continuing in
this way, adapting measurement bases to earlier measurement
results, we get \[ \ldots X_2^{m_2}Z_2^{n_2}X_1^{m_1}Z_1^{n_1}{\rm
(the\,\, correct\,\, wanted \,\, U)}\ket{\psi}.\] Here
$\ket{\psi}$ is generally a state of many qubits and $X_i$, $Z_i$
are Pauli operations on the $i^{\rm th}$ qubit. The indices $m_i$,
$n_i$ are accumulations (actually bit sums mod 2) of measurement
outcomes.

The same idea applies to the 2-qubit gate $CZ$ too where the
situation is even better. We have the propagation relations:
\[ CZ (Z\otimes I)=(Z\otimes I)CZ \hspace{5mm} CZ (X\otimes I) =
(X\otimes Z) CZ \] (and similarly for $X$ and $Z$ acting on the
second qubit on the LHS's, as $CZ$ is symmetrical). Thus we can
propagate Pauli operators through $CZ$ while keeping $CZ$ the {\em
same} i.e. no basis adaption is required!

\noindent {\bf Remark:}\,\, More generally for any 1-qubit $U$ we
can achieve Pauli propagation by \[ UX=X(XUX) \hspace{5mm}
UZ=Z(ZUZ) \] where we adapt $U$ to change into $XUX$ or $ZUZ$. But
the latter operators look very different from $U$ and we have
(unnecessarily) also preserved the actual identity of the Pauli
operations in this propagation. The actual relations we used above
for $CZ$, $R_x(\theta)$, $R_z(\theta)$ and $W(\theta)$ exploit the
opportunity of allowing the Pauli operations to change while
keeping the adapted operation similar to the original one (e.g.
differing only by a sign in the angle). $\Box$

If $U$ is the total unitary effect of a gate array on a
multi-qubit state $\ket{\psi}$ then using the above methods we are
able to generate a state of the form $\ldots
X_2^{m_2}Z_2^{n_2}X_1^{m_1}Z_1^{n_1} U\ket{\psi}$. In a quantum
computation we finally measure (some qubits of) $U\ket{\psi}$ in
$Z_i$-bases $\{ \ket{0},\ket{1}\}$ and the presence of the Pauli
operations $X_i^{m_i}Z_i^{n_i}$ cause no problems: $Z_i^{n_i}$ has
no effect on the measurement outcomes and $X_i^{m_i}$ requires
only a simple reinterpretation of the output results -- for a
single qubit, if $U\ket{\psi}= a\ket{0}+b\ket{1}$ then
$XU\ket{\psi}=a\ket{1}+b\ket{0}$. Hence we simply need to
reinterpret measurement outcome $k_i$ (0 or 1) as $k_i\oplus m_i$
where $m_i$ is the corresponding $X$ Pauli exponent.

We note a rather curious feature here: these ``final'' $Z$
measurements are {\em never adaptive} (being fixed as $Z$
measurements) so they can always be performed {\em first} before
any of the other measurements have been implemented! i.e. the
output of the computation can be measured before any of the
computation itself has been conducted and the subsequent
measurement outcomes simply serve to alter the interpretation of
those $Z$-measurement outcomes!

\section{Parallelisable computations: Clifford operations}

In our measurement based models, distinct measurements always
apply to {\em disjoint} sets of qubits. Hence they all commute as
quantum operations and if the measurement basis choices were not
adaptive then we could do all the measurements simultaneously, in
parallel. The necessity of adaptive choices arose from the Pauli
propagation relations, but some operations are special in this
regard -- the so-called Clifford operations on $n$-qubits.

Let us introduce the Pauli group $\cp_n$ on $n$ qubits, defined as
the group generated (under multiplication) by $n$-fold tensor
products of $\pm I$, $\pm iI$, $X$ and $Z$. For example $\cp_3$
has elements such as $X\otimes Z\otimes Z$, $-iZ\otimes Y\otimes
I$, $I\otimes I\otimes X$ etc. An $n$-qubit unitary operation $C$
is defined to be a Clifford operation if $C\cp_n C^\dagger =
\cp_n$ i.e. $C\cp_n=\cp_n C$ i.e. for every Pauli operation $P\in
\cp_n$ there is another $P'\in \cp_n$ such that $CP=P'C$. Hence
for any Clifford operation $C$ we can propagate Pauli operations
across $C$ while $C$ stays the same i.e. no adaption is needed
(but the Pauli operation generally changes). For each $n$ the
Clifford operations evidently form a group, called the Clifford
group.

We have already noted that $CZ$ is a 2-qubit Clifford operation.
Also the Hadamard operation has the Clifford property since
$HX=ZH$ and $HZ=XH$. (For any prospective $C$ we need only check
the propagation of $X$ and $Z$ at each qubit to verify the full
Clifford property.) Indeed we can give an explicit description of
the Clifford group on $n$ qubits \cite{gott}. Introduce the
$\pi/4$ phase gate: \[ P_{\pi/4} = \left(\begin{array}{cc} 1 & 0\\
0 & i
\end{array} \right).\] Then we have:

\noindent {\bf Theorem:}\cite{gott}\,\, The Clifford group on $n$
qubits is generated by $Z$, $H$, $P_{\pi/4}$ and $CX$ acting in
all combinations on any of the qubits (i.e. arbitrary arrays of
these gates). $\Box$

Hence any array of Clifford gates $C_k\ldots C_2 C_1 \ket{\psi}$
($\ket{\psi}$ of $n$ qubits) may be implemented in TQC in one
parallel layer of measurements. We get a result of the form
$P_kC_k\ldots P_2C_2P_1C_1 \ket{\psi}$ where $P_i$ are all Pauli
operations on the qubits. Commuting them all out we get
$X_n^{a_n}Z_n^{b_n} \ldots X_1^{a_1}Z_1^{b_1} (C_k\ldots
C_1)\ket{\psi}$ where the indices $a_i$, $b_i$ depend on the
measurement outcomes and the Clifford propagation relations. The
collection of maximally entangled states used in all the
teleportations can also be manufactured in parallel (e.g. apply
$CX$'s to many pairs $\ket{0}\ket{0}$) so the entire quantum
process requires only a {\em constant} amount of quantum parallel
time for any $n$, in contrast to the corresponding gate array
whose depth generally increases with $n$.

However there is a further subtle point here: in addition to the
constant parallel time of the quantum process, the computation of
the Pauli exponents $a_i$, $b_i$ requires a further {\em
classical} computation, which is actually the bitwise sum of
selected measurement outcomes. Hence this classical computation
can be done as a parallel computation of {\em log} depth: to sum
$k$ bits $i_1\ldots i_k$ we first sum all pairs $i_1\oplus i_2$,
$i_3\oplus i_4$, $\ldots$ in parallel and then pairs of the
results etc. At each stage the number of bits is halved so we
require $\log k$ layers to reach the final result. Returning to
the gate array of Clifford operations, it is known that any
Clifford operation on $n$ qubits can be represented as an array of
$O(\log n)$ depth i.e. the same as the total quantum plus
classical depth of the full measurement based implementation. But
the virtue of the measurement based approach is to take a fully
quantum process (in this case, the Clifford gate array of log
depth) and recast it as a quantum process of constant depth plus a
 classical computation (of log depth in our case) i.e. we separate the
original process into a quantum and classical part while suitably
``minimising'' the quantum part. This kind of restructuring is
significant in a scenario where quantum and classical computation
are regarded as separate (or even incomparable) resources (c.f.
further discussion in section \ref{complex} below) and we can ask
interesting questions such as: what is the ``least'' amount of
quantum ``assistance'' needed to supplement classical polynomial
time computation in order to capture the full power of quantum
polynomial time computations?

\noindent {\bf Remark:}\,\, It is known that arrays of quantum
Clifford operations can be classically efficiently simulated. This
is the Knill-Gottesman (KG) theorem \cite{ncbook} which asserts
the following: consider any array of Clifford gates on $n$ qubits,
each initialised in state $\ket{0}$. Let $\cp$ be the probability
distribution resulting from measuring (some of) the output qubits
in the $Z$ basis. Then there is a classical (probabilistic)
computational process which runs in time polynomial in $n$, which
also has $\cp$ as its output distribution.  Furthermore according
to \cite{Aargot} this classical simulation of any Clifford array's
output can always be performed in log depth. In view of this, one
may question the significance of our result above, that any such
Clifford array can be reproduced as a classical (log depth)
process {\em plus} a quantum process of constant depth. However
there is an essential difference in the two representations of the
Clifford array: the KG simulation results in a {\em purely
classical} output (sample of $\cp$) whereas our classical-quantum
separated simulation results in a {\em quantum} state as output
i.e. we are simulating the quantum process itself rather than just
the classical output of some measurement results. Indeed our
result makes an interesting statement about {\em quantum}
properties of Clifford arrays, viz. that the essential quantum
content can be ``compressed down'' into constant quantum depth,
which is not provided by the KG result. $\Box$

\section{The ``one-way'' quantum computer}

We now move on to describe our second measurement based
computational model -- the 1WQC of Raussendorf and Briegel
\cite{rb,rbb} At first sight it looks rather different from TQC
but we will see later that the models are in fact very closely
related.

\begin{figure}[t] \begin{center} \leavevmode\fbox{\parbox[b][35mm][s]{60mm}{
\vfill\footnotesize \includegraphics{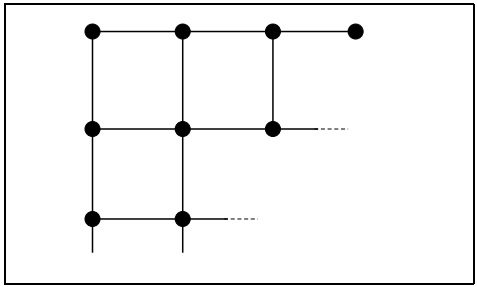}\vfill}}
\caption{\small Construction of the cluster state. The dots
represent a grid of $\ket{+}$ states and a line connecting a pair
of dots represents the application of $CZ$ to the corresponding
pair of qubits.
 }\label{fig3.5}
\end{center} \end{figure}

Consider a rectangular (2 dimensional) grid of $\ket{+}$ states as
shown in figure \ref{fig3.5} . We apply $CZ$ to each nearest
neighbour pair (in horizontal and vertical directions). These
$CZ$'s all commute so for any grid size they can all be applied in
parallel, as a process of constant quantum depth. The resulting
state is an entangled state of many qubits, called a {\em cluster
state}.
We will also use one dimensional cluster states constructed in the
same way, but starting from a one dimensional array of $\ket{+}$
states.

The 1WQC is based on the following facts that will be elaborated
below. Any quantum gate array can be implemented as a pattern of
1-qubit measurements on a (suitably large two dimensional) cluster
state. The only measurements used are in the bases $M_z=\{
\ket{0},\ket{1}\}$ and $M(\theta)=\{ \ket{0}\pm e^{i\theta}\ket{1}
\}$ for some $\theta$. $\theta=0$ corresponds to an $X$ basis
measurement. Measurement outcomes are always labelled 0 or 1 and
they are always uniformly random. As in teleportation, quantum
gates are implemented only up to Pauli corrections $X^aZ^b$ where
$a$ and $b$ depend on the measurement outcomes. (These Pauli
corrections are called bi-product operators in the 1WQC
literature). Hence we'll get the same feature of adaptive
measurements that we saw in TQC. The name ``one-way'' quantum
computer arises from the feature that the initial resource of the
pure cluster state is irreversibly degraded as the computation
proceeds in its layers of measurements.

\begin{figure}[t] \begin{center} \leavevmode\fbox{\parbox[b][40mm][s]{110mm}{
\vfill\footnotesize \includegraphics{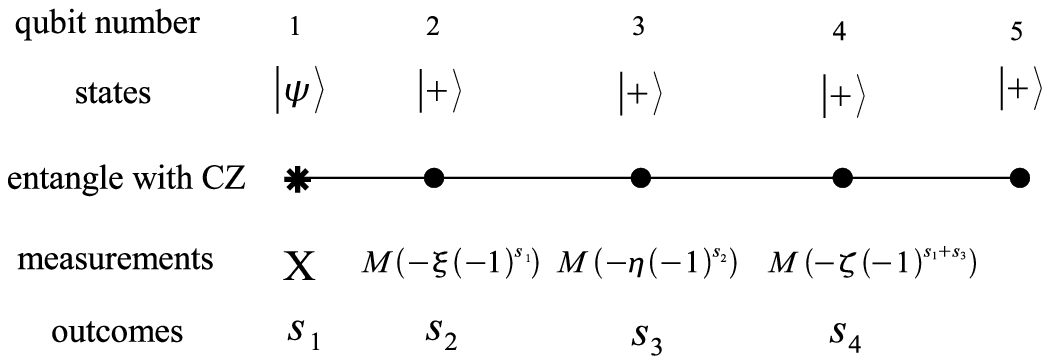}\vfill}}
\caption{\small 1WQC implementation of a 1-qubit unitary $U$ on
$\ket{\psi}$. $\xi$, $\eta$, and $\zeta$ are Euler angles for $U$.
The leftmost qubit, denoted by a star, is set in state
$\ket{\psi}$ and extended by a row of four $\ket{+}$ states
denoted by dots. $CZ$ operations are then applied, denoted by
connecting lines. Next, measurements are applied in the designated
bases with outcomes $s_i$. Hence the measurements must be carried
out adaptively from left to right. As a result of this process the
rightmost (unmeasured) qubit is left in state
$X^{s_2+s_4}Z^{s_1+s_3}U\ket{\psi}$.
 }\label{fig4}
\end{center} \end{figure}

To illustrate these ingredients we give some explicit examples of
measurement patterns for 1-qubit gates (where 1-dimensional
cluster states suffice). Our first example is taken from
\cite{rbb}. We noted previously that any 1-qubit $U$ can be
expressed as \[ U=R_x(\zeta)R_z(\eta)R_x(\xi) \hspace{5mm}
\mbox{for some $\xi,\eta,\zeta$.} \] To apply $U$ to $\ket{\psi}$
by the 1WQC method we start with $\ket{\psi}$ in a line with four
$\ket{+}$ states, as shown in figure \ref{fig4}. (Later we will
see how to eliminate explicit use of $\ket{\psi}$ here, starting
with only $\ket{+}$ states). Entangle all neighbouring pairs with
$CZ$ and subsequently measure qubits 1,2,3 and 4 adaptively in the
bases shown in figure \ref{fig4}. Then it may be shown that qubit
5 acquires the state $X^{s_2+s_4}Z^{s_1+s_3}U\ket{\psi}$, where
$s_i$ is the outcome of the measurement on qubit $i$.
Note that the measurement bases are adaptive and in the above
pattern the measurements must be performed in numerical sequence.

\begin{figure}[t] \begin{center} \leavevmode\fbox{\parbox[b][40mm][s]{60mm}{
\vfill\footnotesize \includegraphics{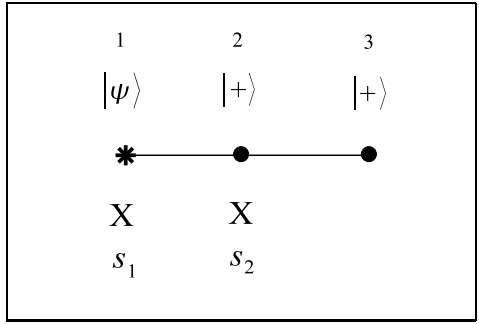}\vfill}}
\caption{\small See caption of figure \ref{fig4} for explanation
of pictorial notations. After the two $X$ measurements, qubit 3 is
left in state $Z^{s_1}X^{s_2}\ket{\psi}$ where $\ket{\psi}$ was
the input state at qubit 1.
 }\label{fig5}
\end{center} \end{figure}

As a second example consider the pattern of figure \ref{fig5}.
Then qubit 3 acquires the state $Z^{s_1}X^{s_2}\ket{\psi}$ so we
have a process which is very similar to teleportation from 1 to 3.
The two $X$ measurements may be done in parallel but we can also
consider this process as a sequence of two identical steps: given
a state $\ket{\psi}$, adjoin $\ket{+}$, entangle with $CZ$ and
then $X$-measure the first qubit, giving an outcome $s$. Qubit 2
is then left in state $X^sH\ket{\psi}$. Thus the two-step chain of
figure \ref{fig5} can be analysed as
$(X^{s_2}H)(X^{s_1}H)\ket{\psi}=
X^{s_2}Z^{s_1}HH\ket{\psi}=X^{s_2}Z^{s_1}\ket{\psi}$ (where we
have used the Pauli propagation relations for $H$).

\begin{figure}[t] \begin{center} \leavevmode\fbox{\parbox[b][40mm][s]{40mm}{
\vfill\footnotesize \includegraphics{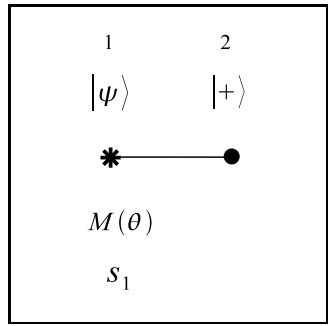}\vfill}}
\caption{\small Effect of a single $M(\theta)$ measurement in
1WQC. Qubit 2 is left in state $X^{s_1}W(-\theta)\ket{\psi}$.
 }\label{fig6}
\end{center} \end{figure}

Figure \ref{fig6} shows a single step operation for the general
measurement basis $M(\theta)$.
Qubit 2 is then left in state $X^{s_1}W(-\theta)\ket{\psi}$ (with
$W(\theta)$ as defined in eq. (\ref{wth})). Indeed if we are not
concerned with issues of parallelisability then any 1-dimensional
measurement pattern may be viewed as a sequential application of
the process of figure \ref{fig6} applied repeatedly. This shows
that the unitary operation $W(\theta)$ plays a fundamental role in
1WQC. For example in figure \ref{fig4} we can reorder all
operations as follows: (entangle 12, measure 1), (entangle 23,
measure 2), etc. This is because the 12-entangling operation and
first measurement both commute with all subsequent entangling
operations and measurements.

In a similar way it is now straightforward to see how gates may be
concatenated. Suppose we wish to apply $U_1$ and then $U_2$ to an
input qubit $\ket{\psi}$. Measurement pattern 1 for $U_1$ has an
output qubit (e.g. qubit 5 in figure \ref{fig4}) which is the
input qubit for the measurement pattern of $U_2$. But all
measurements in pattern 1 commute with all entangling operations
of pattern 2. Hence we can apply all entangling operations (for
both patterns) first, to get a longer single cluster state and
then apply the measurements. Furthermore some measurements in
pattern 2 could even be performed {\em before} those in pattern 1
if their basis choice does not depend on pattern 1 outcomes.

Finally we can eliminate the input state $\ket{\psi}$ from the
above descriptions, to get a formalism based entirely on a
starting state that's a fully standard cluster state (of slightly
longer length): since we can implement any 1-qubit $U$ we can take
our input starting state to be $\ket{+}$ and prefix the desired
process with an initial measurement pattern for a unitary
operation that takes $\ket{+}$ to $\ket{\psi}$.

\begin{figure}[t] \begin{center} \leavevmode\fbox{\parbox[b][50mm][s]{100mm}{
\vfill\footnotesize  \includegraphics{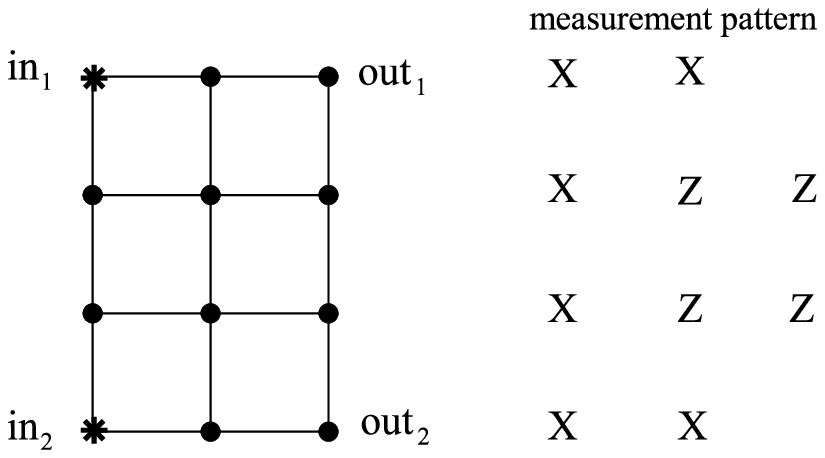}\vfill}}
\caption{\small Application of $CZ$ gate in 1WQC. The 2-qubit
input state $\ket{\psi_{\rm in}}$ is placed at sites labelled
${\rm in}_1$ and ${\rm in}_2$. Dots denote $\ket{+}$ states and
connecting lines denote application of $CZ$ for cluster state
generation. If the measurement pattern shown at the right is
applied at the sites, then only sites ${\rm out}_1$ and ${\rm
out}_2$ remain unmeasured and contain $(P_1\otimes P_2) CZ
\ket{\psi_{\rm in}}$ where $P_1\otimes P_2$ is a Pauli operation
that depends on the measurement outcomes.
 }\label{fig7}
\end{center} \end{figure}

Above we have considered only 1-qubit gates but the formalism may
be generalised (using 2-dimensional cluster grids) to incorporate
2-qubit gates. For universal computation it suffices to be able to
implement just the $CZ$ gate (in addition to 1-qubit gates). An
explicit measurement pattern for $CZ$ is shown in figure
\ref{fig7}. It should also be noted that measurement patterns are
not unique and subject to various approaches for their invention.

\subsection{Role of Z measurements}

The $CZ$ gate requires use of a 2-dimensional grid whereas 1-qubit
gates require only 1-dimensional clusters. Hence measurement
patterns for general gate arrays, implemented on a suitably large
2-dimensional cluster, will generally have some extraneous sites
not used in the measurement patterns. Z-measurements are used to
delete such extraneous sites.

\begin{figure}[t] \begin{center} \leavevmode\fbox{\parbox[b][30mm][s]{60mm}{
\vfill\footnotesize \includegraphics{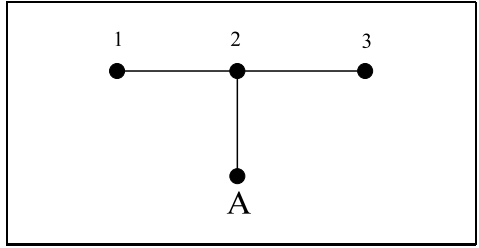} \vfill}}
\caption{\small $Z$ measurements are used to eliminate unwanted
sites such as A.
 }\label{fig8}
\end{center} \end{figure}

To illustrate the principle consider a cluster state with
irregular shape in figure \ref{fig8}.
Suppose we wish to use only the linearly arranged sites 1,2,3 i.e.
we want to delete site A. Consider $Z_A$, a $Z$ measurement at
site A, with outcome $k_A$. To see its effect recall that the
pattern in figure \ref{fig8} is obtained by applying $CZ$
operations to a set of four $\ket{+}$ states, located at the
sites. But measurement $Z_A$ commutes with $CZ_{12}$ and $CZ_{23}$
so starting with the four $\ket{+}$ states we can first do
$CZ_{A2}$ and then $Z_A$ before $CZ_{12}$ and $CZ_{23}$. But
$CZ_{A2}\ket{+}\ket{+}= \ket{0}\ket{+}+\ket{1}\ket{-}$ so after
the $Z_A$ measurement (with outcome $k_A$) the sites 1,2,3 contain
$\ket{+}\ket{(-1)^{k_A}}\ket{+}$ i.e. the effect of the $Z_A$
measurement is to have used the Hadamard basis state
$\ket{(-1)^{k_A}}$ at site 2 instead of the standard $\ket{+}$
(with no site at A) and then entangle as usual. Next consider the
basic 1WQC single step of figure \ref{fig6} with $\ket{-}$ instead
of $\ket{+}$ at site 2: we start with $\ket{\psi}\ket{-}$, apply
$CZ$, then do $M(\theta)$ at qubit 1 for output at qubit 2. But
this is the same as the following: start with $\ket{\psi}\ket{+}$,
then (i) apply $CZ$, (ii) do $M(\theta)$ on qubit 1, then (iii)
perform the unitary operation $I_1\otimes Z_2$. This is identical
to the previous process because (iii) commutes with (i) and (ii)
so we can change the order to (iii), then (i) then (ii) and note
that $Z_2\ket{+}=\ket{-}$. A similar argument applies to each
neighbour of A if there is more than one. In summary, we see that
a $Z_A$ measurement (outcome $k_A$) at a site A adds in an extra
$Z^{k_A}$ Pauli correction at all neighbouring sites of A, in
addition to the usual Pauli operations arising from measurement
patterns in a standard cluster state that had site A absent from
the start.

$Z$ measurements also have a second role: as in TQC they are
applied to a final state of a computation to produce the classical
output results. So just as in TQC we have the curious feature that
these ``final'' $Z$ measurements are always non-adaptive and can
be applied first, before any of the computation itself has been
implemented!

To briefly summarise the 1WQC model, we have seen that any quantum
gate array can be translated into a pattern of 1-qubit
measurements on a suitably large 2-dimensional cluster state.
Choices of measurement bases are generally adaptive i.e. possibly
depending on previous measurement outcomes. Thus the 1-qubit
measurements are organised into layers and the measurements within
each layer can be done simultaneously in parallel. The output
qubits (always measured in the $Z$ basis) can always be done in
the first layer. This temporal sequencing ``cuts across'' the
temporal sequence of the original gate array -- all gates are
generally done ``partially'' in each layer and simultaneously
built up as the layers accumulate. Just as in TQC, arrays of
Clifford operations can always be fully implemented with only {\em
one} layer of measurements.

\section{Further features of 1WQC}

\subsection{A further parallelisability result}

Any polynomial sized quantum gate array can be implemented in 1WQC
using at most a polynomial number of measurement layers (c.f.
section \ref{complex} later). Also we have seen that any array of
Clifford operations can be implemented with just one layer of
measurements. This suggests the following interesting question:
which classes of quantum gate arrays can be implemented in 1WQC
with constraints on the number of measurement layers e.g. using 2
or 3 or a logarithmic number of layers? Does the latter include
all polynomial time quantum computation? Although very little is
known about such questions, we have the following result of
Raussendorf and Briegel \cite{rbb,rbjqc}. Its proof depends on
more subtle properties of Pauli propagation relations for
particular operations.

\noindent {\bf Theorem:}\,\, Any gate array using gates from the
set $\{ CX, R_x(\theta)\,\, {\rm all} \,\, \theta \}$ or from the
set $\{ CX, R_z(\theta)\,\, {\rm all} \,\, \theta \}$ can be
implemented with just two measurement layers.

\noindent {\bf Remark:} Neither of these sets is believed to be
universal although it is known that $CX$ with all $y$-rotations
{\em is} universal \cite{shi}.

\begin{figure}[t] \begin{center} \leavevmode\fbox{\parbox[b][40mm][s]{60mm}{
\vfill\footnotesize \includegraphics{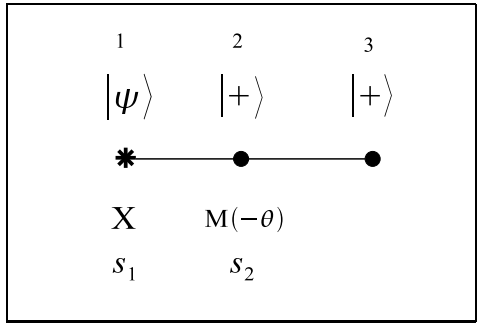}\vfill}}
\caption{\small The process above leaves qubit 3 in state
$X^{s_2}Z^{s_1}R_x((-1)^{s_1}\theta)\ket{\psi}$. Thus to implement
a given $x$ rotation the sign of the angle $\theta$ is adaptive,
depending on the measurement outcome $s_1$.
 }\label{fig14}
\end{center} \end{figure}

\noindent {\bf Proof of theorem:}\,\, We give a proof for
$x$-rotations. (The case of $z$-rotations is similar). We use the
following three (easily verified) facts: (i) any single
$M(\theta)$ measurement (e.g. as shown in figure \ref{fig6})
generates only $I$ or $X$ Pauli corrections. They may become $Z$'s
only after propagation through other gates. (ii) when Pauli
operations are commuted across the Clifford operation $CX$, $X$
propagates only to $X$'s (and $Z$ propagates only to $Z$'s)
although the Pauli operator may spread from one qubit onto two
qubits. (iii) The $R_x$ gate may be implemented in 1WQC using the
measurement pattern shown in figure \ref{fig14}.

Now consider any gate array $\ldots G_3G_2G_1$ where each $G_i$ is
a $CX$ or an $x$-rotation gate. Set up its corresponding
measurement pattern (using figure \ref{fig14} for each
$x$-rotation gate). $CX$ is a Clifford operation and involves no
adaptive measurements. In the first measurement layer we perform
all $CX$ pattern measurements and all $X$ measurements of the
$R_x$ gates (i.e. all qubit 1 measurements in figure \ref{fig14}).
This produces some extraneous Pauli operations and leaves only the
$M(-\theta)$ nodes unmeasured (i.e. all qubit 2's in figure
\ref{fig14}). Next commute all these Pauli's out to the left hand
end of the gate array. This commutation leaves all $CX$'s
unchanged (as $CX$ is Clifford) but some $x$-rotation angles
acquire an unwanted minus sign (c.f. propagation relations for
$x$-rotations given previously). Thus reset these altered angle
signs so that the $M(-\theta)$ measurements again give the correct
designated gates i.e. make the adaptive choice of bases for the
next measurement layer. Finally in the second layer, perform all
the $M(-\theta)$ measurements. We again get some further
extraneous Pauli operations generated. By (i) we get only $I$ or
$X$, but these can be harmlessly commuted out to the left since
$X$ commutes with $x$-rotations and $CX$ preserves $X$'s in its
propagation relations (i.e. no $Z$'s are generated). $\Box$

\subsection{Non-universality of one-dimensional clusters}

We have seen that 1WQC with two-dimensional cluster states is
universal for quantum computation. Nielsen and Doherty
\cite{nielchilds} have shown that any 1WQC process based on only
1-dimensional cluster states can be simulated classically
efficiently i.e. in polynomial time in the number of qubits. Thus
the universality of any such model would imply that quantum
computation is no more powerful than classical computation.

\begin{figure}[t]\begin{center} \leavevmode\fbox{\parbox[b][50mm][s]{100mm}{
\vfill\footnotesize \includegraphics{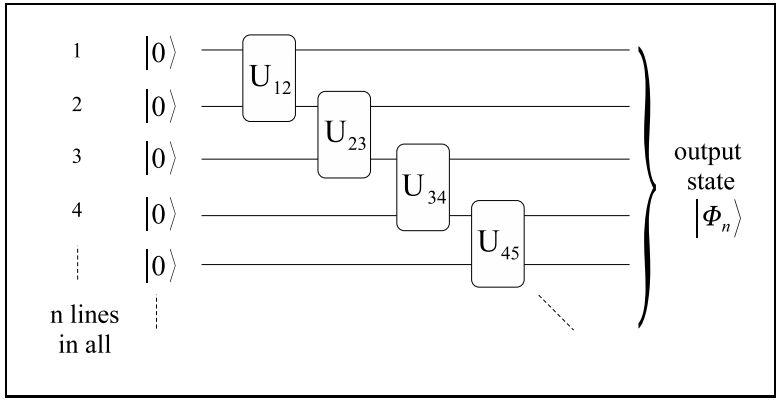}\vfill}}
\caption{\small A generalised 1-dimensional cluster state
$\ket{\Phi_n}$. Starting with a row of $n$ qubits each in state
$\ket{0}$ we sequentially apply $U_{i,i+1}$ to qubits $i,i+1$ as
shown in the above gate array having a ladder-like structure.
 }\label{fig15} \end{center}
\end{figure}

The argument may be paraphrased as follows (ignoring technical
issues of precision of the simulation). Consider any state of the
form shown in figure \ref{fig15}.
(The standard cluster state is obtained by choosing each
$U_{i,i+1}$ to be $CZ(H\otimes H)$). Consider now any sequence of
1-qubit measurements such that for later measurements, the choice
of measurement basis and even the choice of qubit used, may both
depend on outcomes of earlier measurements. Then the whole process
may be classically simulated in polynomial (in $n$) time i.e. the
resulting probability distribution of outcomes may be sampled by
classical means in polynomial time.

The proof runs as follows. At a general stage suppose measurements
on lines $a,b, \ldots ,p,q$ have already been simulated and the
chosen sample outcomes are $a',b', \ldots ,p',q'$ respectively.
Even though these measurements may have been chosen adaptively,
once the outcomes have been specified  (as $a',b', \ldots ,p',q'$)
we can perform the measurements in any order we wish to compute
the joint probability $P(a',b', \ldots ,p',q')$ of the designated
outcome string. Now suppose the next measurement is on line $k$.
To sample its outcome distribution we need to know
\[ P(k'| a',b', \ldots ,p',q') = \frac{P(a',b', \ldots ,k',\ldots
,p',q')}{P(a',b', \ldots ,p',q')}. \] Here $a',b', \ldots ,p',q'$
have their fixed values and we take $k'=0,1$ separately. Having
computed this probability we sample $k'$ to give a definite value,
and continue in the same way for the next measurement on some
line, $l$ say.

To show that this whole process is efficient, we need only show
that $P(a',b', \ldots ,p',q')$ can be computed in poly$(n)$ time
for any chosen set of values $a',b', \ldots ,p',q'$, on any chosen
set of lines. Without loss of generality suppose these lines are
listed in increasing order of occurrence from line 1 at the top.
Note that we can regard a measurement on any line, $a$ say, as
occurring immediately after $U_{a,a+1}$ and before $U_{a+1,a+2}$.
Let $\rho_m$ denote the reduced state of any line $m$ at the
position in between the application of $U_{m-1,m}$ and
$U_{m,m+1}$, and let $\sigma_{m,m+1}$ denote the application of
$U_{m,m+1}$ to $\rho_m \otimes \proj{0}$.

Starting at the top we compute $\sigma_{12}$ as $U_{12}$ on
$\ket{0}\ket{0}$. Then compute $\rho_2$ by partial trace, tracing
out system 1 from $\sigma_{12}$. Then compute $\sigma_{23}$ as
$U_{23}$ on $\rho_2\otimes \proj{0}$ etc. continuing until
$\sigma_{a,a+1}$ where the first measurement occurs. At this point
we do not compute $\rho_{a+1}$ as the partial trace above but
instead, apply to $\sigma_{a,a+1}$ the projector corresponding to
measurement outcome $a'$, obtaining a subnormalised state
$\rho_{a+1}$ on line $a+1$. In fact $\tr \rho_{a+1}$ is the
probability of getting outcome $a'$ in the $a$-line measurement.
We continue in this way computing the reduced states of the lines
successively (using the measurement projector at any measured line
and partial trace at any unmeasured line) until we have applied
the last ($q'$) measurement projector. The trace of this final
resulting state is then $P(a',b', \ldots ,p',q')$. At each stage
of this calculation we need to hold the state of at most two lines
(i.e. not $n$ lines with its exponentially large description) and
we pass through the ladder $n$ times, once for each successive
measurement. Hence the whole calculation is completed in time
polynomial in $n$. This feature of the calculation is a
consequence of the special ladder-like structure of figure
\ref{fig15} and it does not apply to general gate arrays.

\section{Relationships between the TQC and 1WQC models}

The two models have several similarities -- both are based on
measurements as computational steps and both have the awkward
feature of supplementing desired gates with unwanted Pauli
operations. But there are also some essential differences: TQC
uses (Bell) measurements on 2 or more qubits whereas 1WQC uses
only 1-qubit measurements. 1WQC starts with a cluster state having
multi-partite entanglement across {\em all} the qubits whereas TQC
is based on a state comprising only {\em bipartite} entangled
pairs.

The models can be related in several different ways. We will
discuss three of them. The relationships serve to improve our
understanding of 1WQC and its measurement patterns. For TQC the
relation between the measurement and the desired gate is already
transparent.

\begin{figure}[t] \begin{center} \leavevmode\fbox{\parbox[b][30mm][s]{100mm}{
\vfill\footnotesize \includegraphics{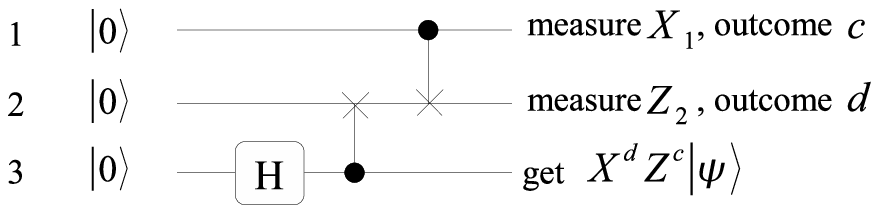}\vfill}}
\caption{\small Teleportation as entangling operations and 1-qubit
measurements. The diagram is read from left to right. The vertical
lines denote $CX$ operations with control and target marked as
$\bullet$ and X respectively. Note that the two $CX$ operations,
with the shown opposite orientations, commute. Alternatively we
could have put $H$ on line 2 and the first $CX$ the other way up,
but then the two $CX$'s would not commute.
 }\label{fig9}
\end{center} \end{figure}

Our first way of relating TQC and 1WQC was proposed by Aliferis
and Leung\cite{AL}. The basic idea is to identify suitable {\em
pairs} of consecutive 1-qubit measurements in 1WQC with a Bell
measurement of a teleportation. This approach is strongly
suggested by patterns such as the one in figure \ref{fig5}. With
reference to the basic teleportation scheme in figure \ref{fig1}
we note the following. $CX$ transforms Bell states into product
states: \[ CX \ket{B_{ij}}= \ket{(-1)^j}\ket{i} \] (where the
first ket is a $\ket{\pm}$ state according to the given sign).
Hence the Bell measurement on 12 can be performed by the
entangling operation $CX$ followed by the 1-qubit measurements
$X_1$ and $Z_2$. Similarly the $\ket{B_{00}}$ state of 23 is
$CX\ket{+}_2\ket{0}_3$ which we can alternatively write as $CX
\ket{+}_3\ket{0}_2$ (where subscripts denote the qubit number and
we adopt the notation for the asymmetrical $CX$ operation that the
first qubit listed is the control qubit). Thus teleportation fully
decomposes into entangling operations and 1-qubit measurements as
shown in figure \ref{fig9}.
So, in view of figure \ref{fig9} teleportation can be interpreted
as: start with $\ket{\psi}_1\ket{0}_2\ket{+}_3$, entangle suitably
with $CX$'s, then do $X_1$ and $Z_2$ measurements, which is
structurally just like the 1WQC paradigm. But we have a slight
mismatch in choice of primitives: TQC uses $CX$ and $\ket{B_{00}}$
states whereas 1WQC is based on $CZ$ and
$\ket{H}=CZ\ket{+}\ket{+}$ states, so the correspondence involves
a sprinkling of Hadamard operations to interconvert these
ingredients.

Extending this idea, we find that other pairs of of measurements
in 1WQC (such as $X_1M(\theta)_2$ and $M_1(\theta)X_2$) can be
interpreted as rotated Bell measurements, but only special pairs
of such consecutive 1-qubit measurements can be fused together to
form Bell measurements. We refer to \cite{AL} for further details
that we will not need here. Although we are able to reconstruct
rotated Bell measurements for the 1WQC implementation of a full
universal set of gates, this interpretation of 1WQC has the
drawback that single 1-qubit measurements individually cannot be
interpreted in terms of TQC.

Our second relationship between TQC and 1WQC, proposed by Childs
et al.\cite{CLN} and Jorrand et al.\cite{JP}, is a further
development of the ideas in figure \ref{fig9}. As noted in the
caption, the two $CX$ operations commute. Also the $X_1$
measurement commutes with all subsequent operations on 23. Thus we
can change the order of actions to: (entangle 12, measure $X_1$),
then (entangle 23, measure $Z_2$), obtaining a sequence of two
operations of the same form viz. (entangle 12, measure 1) to
obtain a state at 2. As noted in figure \ref{fig6}, it is exactly
this kind of operation that drives 1WQC too, so we can regard it
as a common fundamental primitive underlying both models (and
sometimes called ``one-bit teleportation'').

A slight awkwardness in figure \ref{fig9} is the lack of
uniformity of actions: the $CX$'s act in opposite orientations
(necessary for commutativity) and we have a single $H$ gate as
well. But this can be easily remedied: instead of the usual Bell
state based teleportation scheme we consider teleportation with
maximally entangled state $\ket{H} = CZ\ket{+}\ket{+}$ at 23 and
its associated Bell measurement on 12 given by the basis
\begin{equation}\label{bellH} \{ I\otimes I\ket{H}, X\otimes I
\ket{H}, Z\otimes I\ket{H}, XZ\otimes I\ket{H} \}. \end{equation}
Then note that $CZ$ maps this basis to $\{
\ket{+}\ket{+},\ket{+}\ket{-},\ket{-}\ket{-}, \ket{-}\ket{+} \}$
so the Bell measurement is equivalent to applying $CZ$ and then
measuring $X_1$ and $X_2$. Also unlike $CX$, $CZ$ is symmetrical
so the picture as in figure \ref{fig9} for {\em this}
teleportation process is fully uniform. The two ``one-bit
teleportations'' are now identical, in fact corresponding exactly
to the process in figure \ref{fig5}, implemented sequentially.

\subsection{Matrix product state relationship of TQC and 1WQC}

Our third and most remarkable connection between TQC and 1WQC,
proposed by Verstraete and Cirac\cite{vc}, is based on the
formalism of so-called valence bond solids or matrix product
states. In this correspondence each single 1-qubit measurement of
1WQC will be interpreted in terms of a full single teleportation.

\begin{figure}[t] \begin{center} \leavevmode\fbox{\parbox[b][50mm][s]{70mm}{
\vfill\footnotesize \includegraphics{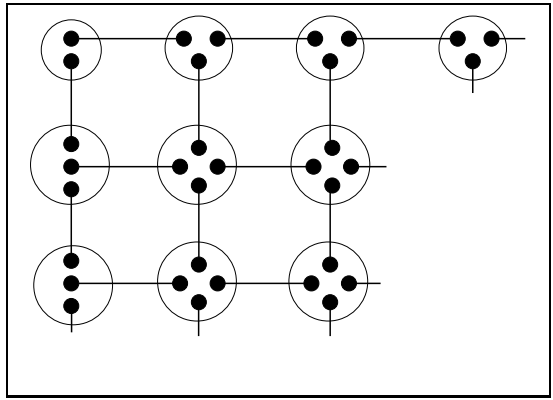}\vfill}}
\caption{\small Valence bond solid or matrix product state for the
cluster state. Each ``valence bond'' line denotes the maximally
entangled state $\ket{H}$ of two qubits. Each site (circled) has 4
or 3 or 2 qubits. At each site we consider the two dimensional
subspace spanned by the two kets of ``all zeroes'' and ``all
ones''.
 }\label{fig10}
\end{center} \end{figure}

Consider a 2-dimensional grid of states $\ket{H}=CZ\ket{+}\ket{+}$
as shown in figure \ref{fig10}. Let $\ket{\rm grid}$ denote the
total state of all the qubits.
At each site in the figure consider the two dimensional subspace
spanned by $\{ \ket{00\ldots 0}, \ket{11\ldots 1} \}$ and the
associated projector, renaming these two basis states as
$\ket{\tilde{0}}$ and $\ket{\tilde{1}}$:
\[ \Pi= \ket{\tilde{0}}\bra{00\ldots
0}+\ket{\tilde{1}}\bra{11\ldots 1}. \] Applying $\Pi$ to $\ket{\rm
grid}$ we obtain a state with a single qubit at each site (and
subnormalised because of the projection).

\noindent {\bf Lemma 3:}\,\, The multi-qubit state $\Pi\ket{\rm
grid}$ (after normalisation) is precisely the 1WQC cluster state.

\begin{figure}[t] \begin{center} \leavevmode\fbox{\parbox[b][30mm][s]{60mm}{
\vfill\footnotesize \includegraphics{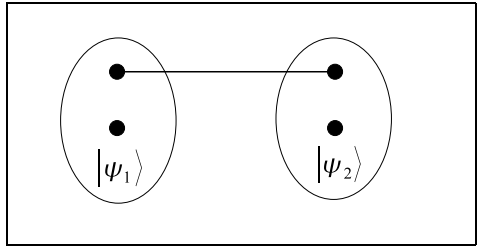}\vfill}}
\caption{\small Dots denote qubits and the connecting line denotes
the maximally entangled state $\ket{H}$. If we project each
circled site to span$\{ \ket{00}, \ket{11} \}$ then the resulting
state (after normalisation) is $CZ\ket{\psi_1}\ket{\psi_2}$.
 }\label{fig11}
\end{center} \end{figure}

\begin{figure}[t] \begin{center} \leavevmode\fbox{\parbox[b][20mm][s]{100mm}{
\vfill\footnotesize \includegraphics{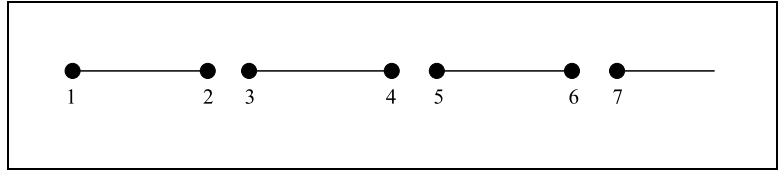}\vfill}}
\caption{\small One dimensional valence bond solid.
 }\label{fig12}
\end{center} \end{figure}

\noindent {\bf Proof:}\,\, We first note the fact of figure
\ref{fig11}.
Now consider a 1-dimensional grid of $\ket{H}$ states as in figure
\ref{fig12}.
Apply $\Pi$ at each node. Each bond is already $CZ\ket{+}\ket{+}$
by definition. By the fact in figure \ref{fig11} the projections
at sites 23 and 45 simply serve to apply $CZ$ between qubits 2 and
5. Hence the whole projected state is just $CZ$ applied to all
connecting pairs in $\ket{+}\ket{+} \ldots \ket{+}$ i.e. the
1-dimensional cluster state. This argument easily generalises to
the 2-dimensional geometry of figure \ref{fig10}.$\Box$

Next consider using $\ket{H}$ states for TQC via application of
rotated versions of the associated basic Bell measurement eq.
(\ref{bellH}). For clarity of the essential idea, consider the
1-dimensional case of figure \ref{fig12}. Let us calculate the
rotated Bell basis corresponding to the 1-qubit gate
\[ W(-\theta)= \frac{1}{\sqrt{2}}\left( \begin{array}{cc} 1 &
e^{-i\theta} \\ 1 & -e^{-i\theta} \end{array} \right). \] The
basis is given by \[ \ket{a} = (W(-\theta)^\dagger \sigma_a)
\otimes I\ket{H} \] where $a=0,1,2,3$ and $\sigma_0=I$,
$\sigma_1=X$, $\sigma_2=Z$, $\sigma_3=XZ$. A direct calculation
gives the first two states as \[ \begin{array}{c}
\ket{a=0}=\frac{1}{\sqrt{2}} (\ket{00}+e^{i\theta}\ket{11}) \\
\ket{a=1}=\frac{1}{\sqrt{2}}
(\ket{00}-e^{i\theta}\ket{11})\end{array} \] and the remaining two
span the orthogonal complement of span$\{ \ket{00},\ket{11} \}$ at
the site. Thus remarkably these first two Bell states lie within
the 1-qubit subspace determined by the $\Pi$-projection.
Furthermore this part of the Bell measurement corresponds
precisely to the measurement basis $M(\theta)= \{
\ket{\tilde{0}}\pm e^{i\theta}\ket{\tilde{1}} \}$ on the projected
site i.e. on the cluster state. (Note that for other more general
1-qubit gates $U$, the corresponding rotated Bell basis states do
not generally lie in a simple way relative to the $\Pi$-projected
subspaces.)

Stated otherwise, $M(\theta)$ measurements on the cluster state
(which is the basic ingredient of 1WQC, c.f. figure \ref{fig6})
can be thought of as teleportations in the TQC formalism, where
the teleportations always produce one of the $a=0,1$ outcomes (and
not $a=2,3$) i.e. teleportations that have been ``cut down'' by
the $\Pi$ projection. In a similar way all other 1WQC ingredients
(viz. $Z$ measurements and the $CZ$ measurement pattern) can be
seen as descendants under the $\Pi$ projection of teleportations
on a valence bond grid state $\ket{\rm grid}$. We omit further
details which may be found in \cite{vc}.

\section{Measurement based models and computational complexity}
\label{complex}

The gate array model of quantum computation provides a transparent
formalism for the theoretical study of quantum computation and its
computational complexity features compared to classical
computation. So why should we bother with further exotic models
such as the measurement based models? Indeed our measurement based
models are readily seen to be polynomial time equivalent to the
gate array model i.e. each model can simulate the other with only
a polynomial (i.e. modest) overhead of resources (number of qubits
and computational steps). To see this first recall that the
standard gate array model (allowing measurements only at the end
and only in the $Z$ basis) can be easily generalised to allow
measurements along the way with subsequent choices of further
gates and measurements being allowed to depend on earlier
measurement outcomes. Indeed consider a measurement in a basis $\{
U\ket{0},U\ket{1} \}$ on a qubit B applied during the course of a
gate array process. To regain the standard gate array paradigm,
for each such measurement we adjoin an extra ancillary qubit A,
initially in state $\ket{0}$ and replace the measurement by the
following: apply $U^\dagger$ to B and the apply $CX$ to qubits BA.
This simulates a coherent representation of the measurement in
which qubit A plays the role of a pointer system. Subsequent gates
that depend on the measurement outcome are replaced by a
corresponding controlled operation, controlled by the state of A
(written in the $Z$ basis). In this way we purge all intermediate
measurements from the body of the array and a measurement of each
ancilla in the $Z$ basis at the end results in a standard gate
array process which is equivalent to the given non-standard one.

Using the above technique any 1WQC process is easily converted
into an equivalent (standard) gate array process. We first build
the required cluster state using an array of $CZ$ gates acting on
$\ket{+}=H\ket{0}$ states and introduce an ancilla A for each
1-qubit 1WQC measurement. For each $M(\theta)$ measurement we
introduce an extra gate $W^\dagger(\theta)$ which transforms the
$M(\theta)$ basis to the standard basis. The overhead in number of
qubits and gates in this simulation is at most linear.

Conversely given any gate array (based say on one of our
previously considered universal sets of gates) we have seen how it
can be translated into a measurement pattern on a suitably large
cluster state. If $K$ is the largest size of the 1WQC measurement
pattern for any gate in our universal set then the number of
qubits and computational steps increases by at most a factor of
$K$ i.e. the resource overhead is again linear.

Polynomial time equivalence of computational models is important
in computational complexity theory because such models have the
same class of polynomial time computations. But polynomial time
equivalence does not preserve more subtle structural features of
computations, such as parallelisability. Indeed already in the
context of classical computation it is well known that the (one
tape) turing machine model is polynomial time equivalent to the
(classical) gate array model yet the turing machine model does not
even have a natural notion of parallalisability at all, whereas
the gate array model does! (i.e. doing gates simultaneously in
parallel).

In contrast to the quantum gate array model, the formalism of
measurement based models offers new perspectives for
parallelisability issues. We have already noted the fundamental
feature that measurements on different subsystems of an entangled
state always commute so long as the choice of measurement is not
adaptive i.e. not dependent on the outcome of another measurement.
We have seen examples of processes which are inherently sequential
for gate arrays (e.g. sequences of Clifford gates) that become
parallelisable in the measurement based models.

The measurement based models have a further novel feature: they
provide a natural formalism for separating a quantum algorithm
into  ``classical parts and quantum parts''. In contrast, in the
gate array model every computational step is viewed as being
quantum. The notion of classical-quantum separation becomes more
compelling when we consider say, Shor's algorithm in its full
totality, including the significant amount of non-trivial
classical post-processing of measurement results needed to reach
the final answer. It seems inappropriate to view this
post-processing as a quantum process (albeit one that maintains
the computational basis)!

In measurement based computation the quantum parts of the
algorithm are the quantum measurements done in parallel layers and
the interspersed classical parts correspond to the adaptive
choices of measurement bases, determined by classical computations
on the previous layers' measurement outcomes. We may generalise
this formalism in the following way: we allow (adaptively chosen)
unitary gates as well as measurements within the quantum parts. We
allow quantum layers to have only depth 1 (so a depth $K$ quantum
process is regarded as $K$ layers with no interspersed classical
computations) whereas classical layers can have any depth i.e. we
are less concerned about controlling their structure.

In this formalism any quantum computation is viewed as a sequence
of classical and quantum layers. The total quantum state is passed
from one quantum layer to the next and the quantum actions carried
out in the next layer are determined by classical computations on
measurement outcomes from previous layers.

Any polynomial time quantum computation (say in the gate array
model) can clearly be implemented with a polynomial number of
quantum layers (and no interspersed classical layers) but the
above formalism suggests a novel structural conjecture:

\noindent {\bf Conjecture:} \,\, Any polynomial time quantum
algorithm can be implemented with only $O(\log n)$ quantum layers
interspersed with polynomial time classical computations. $\Box$

This conjecture, asserting an exponential reduction in the
essential ``quantum content'' of any quantum algorithm, has no
analogue in classical complexity theory (where there is no notion
of classical-quantum separation). Intuitively we are conjecturing
that polynomial time classical computation needs relatively little
``quantum assistance'' to achieve the full power of polynomial
time quantum computation. Although the conjecture remains unproven
in general, we note that Cleve and Watrous \cite{cw00} have shown
that it holds true for Shor's algorithm.

\section{Acknowledgements}
These notes were developed in the course of presentations of the
subject matter at a series of summer schools and workshops: the
CNRS summer school on quantum logic and communication, Corsica,
August 2004, the workshop on quantum information and computation,
Newton Institute, Cambridge UK, August -- December 2004, the
NATO-ASI summer school on quantum computation and quantum
information, Crete, May 2005 and the JST summer school on quantum
information, Kochi, Japan, August 2005. Thanks to Sean Clark and
Noah Linden for discussions of the material and to Sean Clark for
providing the figures. This work was partially supported by the EU
project RESQ-IST-2001-37559 and the UK EPSRC Interdisciplinary
Research Collaboration on Quantum Information Processing.


\end{document}